\documentclass[10pt, conference]{IEEEtran}
\usepackage{epsfig,amsfonts,subfigure,color,amsbsy}
\usepackage[nolist]{acronym}
\usepackage{graphicx,cite,amssymb,amsthm,bm}
\usepackage{amsmath}
\usepackage{mathrsfs}
\usepackage{multirow}
\usepackage{bigstrut}
\usepackage{psfrag}
\usepackage{stackengine}
\usepackage{xcolor}
\usepackage{mathtools}
\usepackage{tabularx,array}
\usepackage{mathbbol}
\usepackage{lipsum}
\usepackage[sans]{dsfont}
\usepackage{array, makecell}
\usepackage{float}
\usepackage{balance}
\usepackage{verbatim}
\usepackage{mathrsfs}
\usepackage{tikz}
\usepackage{pgfplots}
\usepackage{scalerel}
\usepackage{multirow}
\usepackage{graphicx}
\usepackage{pgfplots}
\usepackage{tikz,bm,scalerel}
\usepackage{amssymb}
\usepackage{amsfonts}
\usepackage{mathrsfs}
\usepackage{fancyref}
\usepackage{textcomp}
\usepackage{color}
\usepackage{bbm}
\usepackage{xspace}

\pgfplotsset{compat=1.18}

\usepackage{tikz}
\usetikzlibrary{shapes,arrows}
\usetikzlibrary{calc}

\newtheorem{definition}{Definition}
\newtheorem{lemma}{Lemma}

\renewcommand{\mathbf}{\bm}

\newcommand{\dec}{\text{\rm \texttt{DEC}}}

\newcommand{\argmax}[1]{\underset{#1}{\mathrm{arg \, max}}\,}
\DeclareMathOperator*{\Span}{span}

\newcommand{\pNull}{\epsilon_{\scaleto{\mathsf{0}}{3.5pt}}}
\newcommand{\pOne}{\epsilon_{\scaleto{\mathsf{1}}{3.5pt}}}
\newcommand{\pTwo}{\epsilon_{\scaleto{\mathsf{2}}{3.5pt}}}

\newcommand{\Hnull}{\mathcal{H}_{\scaleto{0}{3.5pt}}}
\newcommand{\Hone}{\mathcal{H}_{\scaleto{1}{3.5pt}}}

\newcommand{\code}{\mathcal{C}}

\newcommand*{\fancyrefapplabelprefix}{app}		
\newcommand*{\fancyrefthmlabelprefix}{thm}		
\newcommand*{\fancyreflemlabelprefix}{lem}		
\newcommand*{\fancyrefcorlabelprefix}{cor}		
\newcommand*{\fancyrefdeflabelprefix}{def}		
\newcommand*{\fancyrefproplabelprefix}{prop}		
\newcommand*{\fancyrefexelabelprefix}{exe}		
\newcommand*{\fancyrefpropolabelprefix}{propo}		

\usepackage{./style/style_Sauter}
\usepackage{./style/standard-macros}

\IEEEoverridecommandlockouts
\begin{document}


\title{Polar and Convolutional Codes for the Unequal Message Protection Problem}

\author{\IEEEauthorblockN{Alexander Sauter\IEEEauthorrefmark{1}\IEEEauthorrefmark{2}, Riccardo Schiavone\IEEEauthorrefmark{1}, Lucía Balsa Picado\IEEEauthorrefmark{1}\IEEEauthorrefmark{2}, and
Gianluigi Liva\IEEEauthorrefmark{1}}
\IEEEauthorblockA{\IEEEauthorrefmark{1}\textit{German Aerospace Center,
We{\ss}ling, Germany}}
\IEEEauthorblockA{\IEEEauthorrefmark{2}\textit{Institute for Communications Engineering, Technical University of Munich, Munich, Germany}} 
\{alexander.sauter, riccardo.schiavone, gianluigi.liva, lucia.balsapicado\}@dlr.de
\thanks{Alexander Sauter, Riccardo Schiavone, and Gianluigi Liva would like to thank the Federal Ministry of Research, Technology, and Space (BMFTR) for supporting the xG-RIC project as part of the research program Communication Systems “Souverän. Digital. Vernetzt.” (grant number 16KIS2429K).
}} 



 \maketitle

 \IEEEoverridecommandlockouts
 



\begin{abstract}
This paper proposes the design of polar and convolutional coset codes for the \ac{UMP} in the short blocklength regime, to overcome the rate loss introduced by preamble-based solutions. After providing conditions to ensure message class disjointness, a two-step decoding architecture is proposed: it first identifies the message class via a likelihood ratio test—computable exactly for convolutional codes and approximated for polar codes—and subsequently performs maximum (or near) likelihood decoding among the codewords of the chosen message class. Numerical results show that our construction closely tracks finite-length benchmarks. Specifically, the analyzed CRC-aided polar codes perform comparable to existing polar code approaches, without requiring specific code design, while offering a robust and spectrally efficient solution for \ac{UMP} scenarios.
\end{abstract}



\begin{IEEEkeywords}
unequal message protection, polar codes, convolutional codes, coset codes, likelihood ratio test.
\end{IEEEkeywords}


\begin{acronym}
    \acro{AWGN}{additive white Gaussian noise}
    \acro{ALRT}{approximate likelihood ratio test}    
    \acro{biAWGN}{binary-input additive white Gaussian noise}
    \acro{BBT}{binary balanced tree}
    \acro{B-DMC}{binary-input discrete memoryless channel}
    \acro{BCJR}{Bahl-Cocke-Jelinek-Raviv }
    \acro{BP}{belief propagation}
    \acro{BPSK}{binary phase shift keying}
    \acro{BSC}{binary symmetric channel}
    \acro{BEC}{binary erasure channel}
    \acro{BP}{belief propagation}
    \acro{CA}{CRC-aided}
    \acro{DE}{density evolution}
    \acro{GA}{Gaussian approximation}
    \acro{GLRT}{generalized likelihood ratio test}    
    \acro{GLLRT}{generalized log-likelihood ratio test}
    \acro{i.i.d.}{independent and identically distributed}
    \acro{DMC}{discrete memoryless channel}
    \acro{LDPC}{low-density parity-check}
    \acro{LRT}{likelihood ratio test}
    \acro{LLRT}{log-likelihood ratio test}
    \acro{LLR}{log-likelihood ratio}
    \acro{LHS}{left-hand-side}
    \acro{ML}{maximum likelihood}
    \acro{NA}{normal approximation}
    \acro{r.v.}{random variable}
    \acro{PEP}{pairwise error probability}
    \acro{p.m.f.}{probability mass function}
    \acro{p.d.f.}{probability density function}
    \acro{CCC}{constant composition code}
    \acro{SNR}{signal-to-noise ratio}
    \acro{EXIT}{extrinsic information transfer}
    \acro{PEXIT}{protograph extrinsic information transfer}
    \acro{VN}{variable node}
    \acro{CN}{check node}
    \acro{FER}{frame error rate}
    \acro{MN}{MacKay-Neal}
    \acro{MI}{mutual information}
    \acro{RA}{repeat-accumulate}
    \acro{RHS}{right-hand-side}
    \acro{RV}{random variable}
    \acro{NS}{non-systematic}
    \acro{UER}{undetected error rate}
    \acro{UEP}{unequal error protection}
    \acro{UMP}{unequal message protection}
    \acro{SC}{successive cancellation}
    \acro{GSCL}{generalized successive cancellation list}
    \acro{SCL}{successive cancellation list}
    \acro{TEP}{total error probability}
     \acro{ASCL}{augmented successive cancellation list}
    \acro{CRC}{cyclic redundancy check}
    \acro{RCB}{random coding bound}
    \acro{eMBB}{enhanced mobile broadband}
    \acro{mMTC}{massive machine-type communication}
     \acro{URLLC}{ultra-reliable low-latency communication}
     \acro{RCU}{random coding union}
     \acro{ZTCC}{zero-tail terminated convolutional code}
\end{acronym}

\section{Introduction}\label{sec:intro}

Modern communication systems, like those aimed to provide ultra-reliable low-latency communications \cite{3GPP21}, increasingly rely on the transmission of short information blocks, rather than large-scale packets. This paradigm shift has sparked a renewed interest in the finite-blocklength fundamental limits of channel coding , moving beyond traditional asymptotic analysis to accurately characterize achievable rates. A seminal contribution to this field is the work by Polyanskiy, Poor and Verdú \cite{polyanskiy2010channel}, which established tight performance bounds for the best codes in the short-packet regime. In such scenarios, convolutional and polar codes have emerged as highly competitive solutions, closely approaching these theoretical limits \cite{TV15,gaudio2017performance,cocskun2019efficient,yang2022crc,sui2023crc}.

Within these communication frameworks, different classes of messages often coexist, each characterized by distinct reliability constraints or error probability targets. This paradigm is formally known as \ac{UMP} \cite{Shkel14,Shkel15} (or \textit{message-wise} \ac{UEP} in \cite{Borade09ISIT}). A specific instance of the \ac{UMP} framework is the ``red alert'' problem \cite{Nazer13}, where a single critical message must be received with significantly higher reliability compared to all others.

A common practical approach to handle \ac{UMP} involves the use of physical layer preambles or headers to identify the message class prior to decoding. While effective for long packets---as seen in standards like DVB-S2\cite{ETSIDVBS2}, where different modulation and coding options coexists---this strategy incurs significant rate penalties at short blocklength, since the preamble overhead becomes a non-negligible fraction of the total resources. To overcome this loss, in \cite{Shkel14,Shkel15} the use of coset codes is proposed as alternative and it is analyzed through the lens of finite-length information theory.

This paper investigates the use of coset-based \ac{UMP} constructions that circumvent the need for explicit headers. We derive formal criteria and simple lemmas to ensure the disjointness of the component codebooks, preventing irreducible error floors caused by non-null intersections. Furthermore, we propose a two-step decoding architecture designed to identify the message class and subsequently estimate the transmitted codeword. This approach utilizes a \ac{LRT} that is exactly computable for \acp{ZTCC} via trellis-based recursions. For polar codes \cite{Ari09}, we introduce an approximate version. Our results show that this decoding framework is both simple and robust, offering a practical solution to \ac{UMP} in short-blocklength scenarios.

\section{Preliminaries}\label{sec:preliminaries}
We denote by $(n, k_i)$ the blocklength and the dimension of the code $\code_i$, respectively. For polar codes\cite{Ari09}, we construct codes by concatenating an outer $(n^{\prime}_i, k_i)$ \ac{CRC} code with an inner $(n, n^{\prime}_i)$ polar code. We assume that \ac{SCL} decoding \cite{TV15} for polar codes with list size $L$ is performed. 
For $(n, k_i)$ \acp{ZTCC}, we puncture the output of a rate $1/\mathtt{n}_i$ nonsystematic convolutional encoders of memory $\nu$ to be exactly $n$. The generator polynomials of each \ac{ZTCC} $\code_i$ is defined as $G_i(\mathtt{D})=[g^{(1)}_i(\mathtt{D}), g^{(2)}_i(\mathtt{D}), \dots, g^{(\mathtt{n}_i)}_i(\mathtt{D})]$. More details on the punctured \ac{ZTCC} will be provided in \ref{subsec:ZTCC_UMP}. We consider a \ac{BPSK} transmission under the model 
\begin{equation} \label{eq:channel}
	y_i = x_i + z_i, \quad i=1,2,\ldots,n.
\end{equation}
where $x_i \in \{-1,+1\}$, and the $z_i$ are \ac{i.i.d.} and follow a Gaussian distribution with zero mean and variance $\sigma^2$. For the channel \ac{SNR}, we use $E_s / N_0 = 1/(2\sigma^2)$.  

\section{Unequal Message Protection}\label{sec:UMP}

This section introduces the formal framework for \ac{UMP} codes and discusses their implementation in the finite blocklength regime.

\begin{definition}[\ac{UMP} Code \cite{Shkel15}]
An $(n, \{k_i, \epsilon_i\}_{i=0}^{m-1})$ \ac{UMP} code $\mathcal{C}$ is defined as the union of $m$ disjoint message classes, each one encoded by code $\{\mathcal{C}_i\}_{i=0}^{m-1}$.
\[\code = \bigcup^{m-1}_{i=0} \code_i.\]
Each class contains $M_i = 2^{k_i}$ messages and is characterized by an average decoding error probability constraint $\epsilon_i$. 
The total number of messages in the code is $M = \sum_{i=0}^{m-1} M_i$.
\end{definition}

A straightforward implementation of \ac{UMP} involves the use of a preamble to identify the message class prior to decoding. While this approach is efficient for large $n$---as exemplified by the DVB-S2 standard \cite{ETSIDVBS2}, where physical layer headers signal the modulation and channel code configuration---it incurs in a significant rate penalty when $n$ is small.

In such small blocklength scenarios, the preamble overhead becomes significant, as shown in \cite{Shkel15}, demanding more efficient \ac{UMP} strategies. To this end, we adopt in this paper the coset-based \ac{UMP} constructions proposed in \cite{Shkel15}, and briefly reviewed in \ref{subsec:CosetUMP}, which circumvent the need for explicit headers. Specifically, we focus on the case of $m=2$ message classes, hereafter referred to as \textit{critical} and \textit{non-critical} message classes, associated with the component codes $\mathcal{C}_0$ and $\mathcal{C}_1$, respectively. We refer to the case of $m =2$ as the $2$-UMP code construction.

In particular, we focus on codes in the short blocklength regime i.e. polar and convolutional codes.

The rest of this section is organized as follows: in section~\ref{subsec:BoundUMP}, we adopt the \acl{NA} of~\cite{Shkel15} to assess the performance of the polar and convolutional codes. Section~\ref{subsec:CosetUMP} discussed the construction of \ac{UMP} codes.  

\subsection{Finite-Length Performance Benchmarks}\label{subsec:BoundUMP}

To assess the performance of the proposed \ac{UMP} code constructions, we adopt the \ac{NA} as a benchmark, specifically referring to the \ac{UMP} result in \cite[Th. 19]{Shkel15}. The maximum achievable message size $k_i$ for each class $i \in \{0, \dots, m-1\}$ with an error probability constraint $\epsilon_i$ and blocklength $n$ can be approximated as:
\begin{equation}
    k_i \approx nC\!\left(\!\frac{E_s}{N_0}\!\right)\!-\! \sqrt{nV\!\left(\!\frac{E_s}{N_0}\!\right)} Q^{-1}(\epsilon_i) \!-\! \frac{1}{2} \log_2\! n\! +\! \log_2\! \frac{1}{\lambda_i}
\end{equation}
where $C$ and $V$ denote the \ac{biAWGN} channel capacity and dispersion, respectively, evaluated at \ac{SNR} $E_s/N_0$ and $Q^{-1}(\cdot)$ is the inverse Gaussian $Q$-function. The parameters $\{\lambda_i\}$ are subject to the constraint $\sum_{i=0}^{m-1} \lambda_i = 1$ and are optimized to minimize the $E_s/N_0$ value at which all class of codes satisfy their $(k_i,\epsilon_i)$ requirement.


\subsection{Coset UMP Code Construction}\label{subsec:CosetUMP}

In the following section, we review the definition of coset \ac{UMP} codes, and study the impact of the choice of offset vectors on the floor of the average error probability $\epsilon_i$. 

\begin{definition}[Coset \ac{UMP} Code \cite{Shkel15}]
A coset \ac{UMP} code $\mathcal{C}$ is an \ac{UMP} code where each class code $\mathcal{C}_i$ is an $(n, k_i)$ affine code defined over $\mathbb{F}_2$ as:
\begin{equation}\label{eq:CosetCodeConst}
    \mathcal{C}_i = \left\{ \bm x=\mathbf{u} \mathbf{G}_i + \bmv_i : \bmu \in \mathbb{F}_2^{k_i} \right\},
\end{equation}
where $\mathbf{G}_i \in \mathbb{F}_2^{k_i \times n}$ is the generator matrix of a linear code and $\bmv_i \in \mathbb{F}_2^{n}$ is the offset vector associated with the $i$-th class.
\end{definition}

While the choice of offset vectors $\bm{v}_i$ is often neglected in random coding analysis, it is critical in practical \ac{UMP} implementations. Specifically, an improper selection of $\bm{v}_i$ may lead to non-disjoint codebooks, i.e., $\mathcal{A}=\mathcal{C}_0 \cap \mathcal{C}_1 \neq \emptyset$. A notable example is the all-zero codeword, which is common to all classes if $\bmv_i$ corresponds to the all-zero vector for all $i$. Such intersections introduce ambiguity at the decoder. For instance, if a codeword in the intersection set $\mathcal{A}$ is a priori assigned to class $1$, the error probability for class $0$ becomes:
\begin{equation}
    \epsilon_0 = \frac{|\mathcal{A}|}{M_0} + \frac{M_0 - |\mathcal{A}|}{M_0} \epsilon'_0,
\end{equation}
where $\epsilon'_0$ is the average error probability for the codewords in $\mathcal{C}_0 \setminus \mathcal{C}_1$. In the high \ac{SNR} regime, $\epsilon'_0 \to 0$, leading to an irreducible error floor $\epsilon_0 \to |\mathcal{A}|/M_0$.

Hereafter, we provide ways to measure the cardinality of the intersection set $\mathcal{A}$, in order to construct disjoint codebooks. Firstly, to quantify the impact of codebook overlap, we examine the intersection of two linear codes. This can be efficiently characterized by considering their parity-check representations, because a codeword belongs to both codes if and only if it satisfies the parity-check conditions of both codes.

\begin{lemma}[Dimension of the intersection of two linear codes]
Let $\mathbf{H}_0 \in \mathbb{F}_2^{(n-k_0) \times n}$ and $\mathbf{H}_1 \in \mathbb{F}_2^{(n-k_1) \times n}$ be the parity-check matrices of the linear codes $\mathcal{C}_0$ and $\mathcal{C}_1$, respectively. Then,
\begin{equation}
    \mathcal{C}_0 \cap \mathcal{C}_1 = \left\{ \bmx \in \mathbb{F}_2^n : \begin{bmatrix} \mathbf{H}_0 \\ \mathbf{H}_1 \end{bmatrix} \bmx^T = \bm{0}^T \right\}
\end{equation}
and its dimension is given by
\begin{equation}
    \dim(\mathcal{C}_0 \cap \mathcal{C}_1) = n - \operatorname{rank} \left( \begin{bmatrix} \mathbf{H}_0 \\ \mathbf{H}_1 \end{bmatrix} \right).
\end{equation}
\end{lemma}

The extension to coset codes is straightforward, noting that each codeword in $\code_i$ has the same syndrome
\[\bms_i=(\bm u\bm{G}_i+\bmv_i)\bm{H}_i^T=\bmv_i\bm{H}_i^T.\]

\begin{lemma}[Dimension of the intersection of two coset codes]\label{lemma:cosetcodeintersection}
Let $\mathbf{H}_0 \in \mathbb{F}_2^{(n-k_0) \times n}$ and $\mathbf{H}_1 \in \mathbb{F}_2^{(n-k_1) \times n}$ be the parity-check matrices of the linear codes used to define $\mathcal{C}_0$ and $\mathcal{C}_1$, respectively according to \eqref{eq:CosetCodeConst}. Then,
\begin{equation}\label{eq:intersectioncosetcodes}
    \mathcal{C}_0 \cap \mathcal{C}_1 = \left\{ \bm{x} \in \mathbb{F}_2^n : \begin{bmatrix} \mathbf{H}_0 \\ \mathbf{H}_1 \end{bmatrix} \bmx^T = \begin{bmatrix} \bms_0^T \\ \bms_1^T \end{bmatrix} \right\}
\end{equation}
with $\bms_i=\bmv_i \mathbf{H}_i^T$, and the dimension of the intersection, whenever the system of equation in \eqref{eq:intersectioncosetcodes} admits one or more solutions, is given by:
\begin{equation}
    \dim(\mathcal{C}_0 \cap \mathcal{C}_1) = n - \operatorname{rank} \left( \begin{bmatrix} \mathbf{H}_0,\bms_0^T \\ \mathbf{H}_1,\bms_1^T \end{bmatrix} \right).
\end{equation}
\end{lemma}

In the remainder of the paper, the coset codes are designed by a randomized search of the offsets $\bmv_i$ such that $\mathcal{C}_0 \cap \mathcal{C}_1=\emptyset$ according to Lemma \ref{lemma:cosetcodeintersection}.\\

\section{Decoding of Polar and Convolutional Coset UMP Codes}

In this section we propose a two–step decoding architecture tailored to the 2--UMP coset construction. The decoder first identifies the most likely component codebook, and subsequently performs \ac{ML} decoding within the selected codebook. This section formalizes the optimal \ac{LRT} for the codebook selection, introduces a simplified test, and derives its explicit implementation for \acp{ZTCC}. For polar codes, in the second step \ac{ML} decoding is replaced by \ac{SCL} decoding.




\paragraph{Likelihood Ratio Test}
To identify the message class for the $2$-\ac{UMP} code, we perform a binary hypothesis test. Assuming all codewords in $\code_i$ are equally likely, the likelihood under each hypotheses $\mathcal{H}_i$ is defined as
\begin{align}\label{eq:HypoNullOne}
f_{\bm Y|\mathcal{H}_i}(\bm y | \mathcal{H}_i) = \frac{1}{\lvert \code_i \rvert} \sum_{\bm x\in \code_i} p_{\bm Y|\bm X}(\bm y | \bm x) 
\end{align}
where we refer to $\Hnull$ and $\Hone$ as the hypothesis under which a \textit{critical} and \textit{non-critical} message has been transmitted, respectively. According to the Neyman-Pearson lemma, the optimal test for the two hypotheses $\Hnull$ and $\Hone$ is the \ac{LRT}, and it is given by
\begin{equation} \label{eq:LRT}
    \Lambda_{\mathsf{LRT}}(\bm y) = \frac{f_{\bm Y|\mathcal{H}_0}(\bm y | \Hnull)}{f_{\bm Y|\mathcal{H}_1}(\bm y | \Hone)} \mathrel{\mathop\gtrless\limits^{\Hnull}_{\Hone}}T 
\end{equation}
where $T$ is the decision threshold.
Evaluating \eqref{eq:HypoNullOne} requires summing the likelihood of \emph{all} codewords in $\mathcal{C}_i$. For a general $(n,k)$ binary linear block code, the computational complexity of this marginalization is exponential in $\min(2^k, 2^{n-k})$, and it only becomes tractable for very short codes or particular code families such as terminated convolutional codes, where the trellis representation enables an exact computation \cite{Raghavan1998,Hof2009:ISTCA,Williamson2014:ROVA,Wesel2022:ROVA}.
To avoid the full marginalization in \eqref{eq:HypoNullOne}, we approximate each class–conditional density by the likelihood of its \ac{ML} codeword, yielding the \ac{ALRT}
\begin{equation} \label{eq:ALRT}
    \Lambda_{\mathsf{ALRT}}(\bm y) = \frac{\max_{\bm x \in \code_0}f_{\bm Y|\bm X}(\bm y | \bm x)}{\max_{\bm x \in \code_1}f_{\bm Y|\bm X}(\bm y | \bm x)} \mathrel{\mathop\gtrless\limits^{\Hnull}_{\Hone}}T. 
\end{equation}
We emphasize that the ALRT does \emph{not} eliminate the dependence on the entire codebook: the maximization in \eqref{eq:ALRT} still requires searching over all codewords. The approximation becomes computationally advantageous only when an efficient ML decoding algorithm is available for the specific code family. For other code families, such as polar codes, an ML decoder with manageable complexity is not available, but near ML-decoding is possible, e.g., via \ac{SCL} decoders with sufficiently large list sizes. Therefore, implementing \eqref{eq:ALRT} requires further approximations or tailored decoding strategies, which we detail in the polar decoding subsection.

\paragraph{Error Events}
Let $\hat{\mathcal{H}}\in\{\mathcal{H}_0,\mathcal{H}_1\}$ denote the output of the hypothesis test in the first decoding stage, and let $\hat{\mathbf{x}}$ denote the \ac{ML} codeword estimate produced in the second stage according to
\begin{equation}\label{eq:xhat_clean}
    \hat{\mathbf{x}} = \argmax{{\mathbf{x}\in\hat{\mathcal{C}}}} p_{\mathbf{Y}|\mathbf{X}}(\mathbf{y}\,|\,\mathbf{x})
\end{equation}
where $\hat{\mathcal{C}} = \code_0$ when $\hat{\mathcal{H}}=\mathcal{H}_0$, otherwise $\hat{\mathcal{C}} = \code_1$.

When a \textit{critical} message is transmitted, i.e., under $\mathcal{H}_0$, an error is declared if either the decoder selects the wrong codebook or, despite selecting the correct component, the ML decoder outputs an incorrect codeword. The error probability under $\mathcal{H}_0$ is therefore
\begin{equation}\label{eq:errorNull}
    \pNull = \Prob\left[(\mathcal{\hat{H}} = \Hone) \cup (\mathcal{\hat{H}} = \Hnull \cap \hat{\bm x} \neq \bm x ) \Big| \Hnull\right]. 
\end{equation}
Analogously, when a \textit{non–critical} message is transmitted (hypothesis $\mathcal{H}_1$), the overall error probability is
\begin{equation}\label{eq:errorOne}
    \pOne = \Prob\left[(\mathcal{\hat{H}} = \Hnull) \cup (\mathcal{\hat{H}} = \Hone \cap \hat{\bm x} \neq \bm x ) \Big| \Hone\right]. 
\end{equation}


\subsection{Zero-Tail Terminated Convolutional Codes for UMP}\label{subsec:ZTCC_UMP}

Following \cite{Hof2009:ISTCA}, for ZTCCs, the likelihood functions
$
f_{\bm Y|\mathcal{H}_i}(\bm y | \bm x)$ admit an exact factorization over the trellis, enabling efficient evaluation of the optimal LRT in \eqref{eq:LRT}. Applying the logarithm to the likelihood functions as in \cite{Hof2009:ISTCA} and including additive or multiplicative constants into the threshold test we obtain an alternative formulation based on the metric
\begin{equation}
    \Lambda_{\mathsf{LLRT}}(\bm y) = \ell_{\bm Y|\mathcal{H}_0}(\bm y | \mathcal{H}_0) - \ell_{\bm Y|\mathcal{H}_1}(\bm y | \mathcal{H}_1)
\end{equation}
where, to obtain the $\ell_{\bm Y|\mathcal{H}_i}(\bm y | \mathcal{H}_i)$, we proceed as follows. Denote by
\[
\mu^{j_1\rightarrow j_2}_t
=\frac{1}{\sigma^2}
 \left\langle \mathbf{y}_t,\, \mathbf{x}^{\,j_1\rightarrow j_2}_t \right\rangle
\]
the branch metric associated with the transition from state $j_1$ at trellis section $(t-1)$ to state $j_2$ at trellis section $t$, and denote by $M^{(j)}_t$ the cumulative metric of state $j$ at trellis section $t$.
The forward recursion is 
\begin{equation}
M^{(j_2)}_{t+1}
=\log\!\left(\sum_{j_1\in\mathcal{N}_{j_2}}
     \exp\left( M^{(j_1)}_t
                +\mu^{j_1\rightarrow j_2}_t \right)\right),
\label{eq:forward-log}
\end{equation}
where $\mathcal{N}_{j}$ denotes the predecessor states of $j$ over the trellis, defined as the states at trellis section $(t-1)$ connected to state $j$ at section $t$.
The algorithm is initialized at time $t=0$ for the all-zero state with $M^{(0)}_{0}=0$, and it outputs at the last trellis section $t=k_i+\nu$ of the $i-$th code \[\ell_{\bm Y|\mathcal{H}_i}(\bm y | \mathcal{H}_i)=M^{(0)}_{k_i+\nu}.\]



The ALRT \eqref{eq:ALRT}, instead, corresponds precisely to the ratio of the likelihoods of the choice of ML decoders over the two individual codebooks, which, for ZTCCs, can be efficiently computed via the Viterbi algorithm applied independently to $\mathcal{C}_0$ and $\mathcal{C}_1$.

An advantage of the ALRT is that the ML decoding of the codebooks already provides the candidate solution that follows the hypothesis testing, whereas after the LRT test an instance of Viterbi decoding over the trellis of the selected codebook  is needed in order to identify the estimated codeword $\hat{\bm x}$.

    
\renewcommand{\varepsilon}{\mathsf{E}}

\subsection{Polar Codes for UMP}
Unlike \ac{ZTCC}, for \ac{CA} polar codes an exact evaluation of \eqref{eq:HypoNullOne} is infeasible. Consequently, the decision on the message class is performed via a modification of the \ac{ALRT}, which takes into account the fact that the polar decoders are not \ac{ML} decoders, and that they are \textit{incomplete} decoders.  
Upon receiving the channel output vector $\bm y$, the receiver runs two parallel \ac{SCL} decoders, one for each message class, and denoted by $f_{\dec_i}(\bm y)$. Each \ac{SCL} decoder computes a list $\mathcal{L}_i$ of size $L$ of possible transmitted codewords and it outputs $\hat{\bm{x}}_i$ according to
\begin{equation}
    \hat{\bm{x}}_i\! =\! f_{\dec_i}(\bm y) \! = \!\!
    \begin{cases}
         \argmax{\bm x\in(\mathcal{L}_i\cap \code_i)} p(\bm y \mid \bm x) \!
        & \text{if } \mathcal{L}_i\cap \code_i \neq \emptyset, \\[2.5mm]
        \varepsilon, \!
        & \text{otherwise},
    \end{cases}
\end{equation}
where $\varepsilon$ denotes an erasure event when none of the codewords in $\mathcal{L}_i$ satisfies the CRC parity-check equations. When both $\hat{\bm{x}}_0\neq\varepsilon$ and $\hat{\bm{x}}_1\neq\varepsilon$, the decoder performs the following test
\begin{equation}
    \widetilde{\Lambda}_{\mathsf{ALRT}}(\bm y) = \frac{p(\bm y | \hat{\bm{x}}_0)}{p(\bm y | \hat{\bm{x}}_1)} \mathrel{\mathop\gtrless\limits^{\Hnull}_{\Hone}}T,
\end{equation}
and it outputs $\hat{\bm x}=\hat{\bm{x}}_0$ when it decides for $\Hnull$ otherwise it outputs $\hat{\bm x}=\hat{\bm{x}}_1$. If instead $\hat{\bm{x}}_0=\hat{\bm{x}}_1=\varepsilon$, then the decoder outputs an error flag (erasure), whereas if only one between $\hat{\bm{x}}_0$ or $\hat{\bm{x}}_1$ is an erasure, then $\hat{\bm{x}}$ corresponds to the non-erased value.
Accordingly, an error event occurs every time the decoder outputs a wrong codeword or an erasure.

\section{Results}\label{sec:results}
In this section, we evaluate the performance of the proposed $2$-\ac{UMP} coset constructions using polar and convolutional codes. We consider a \ac{biAWGN} channel and focus on the minimum \ac{SNR} required to satisfy target error probability constraints $(\pNull^\star,\pOne^\star)$ for both message classes simultaneously. We define the \ac{SNR} threshold $(E_s / N_0)^\star$ as the minimum value of $E_s/N_0$ for which the decoder simultaneously satisfies
\[
    \pNull \le \pNull^\star
    \qquad\text{and}\qquad
    \pOne \le \pOne^\star.
\]
For every $2$-\ac{UMP} code, the threshold $T$ of the test is optimized in order to minimize $(E_s / N_0)^\star$. 
The target error probabilities are set to $\epsilon_0^\star = 10^{-5}$ for the \textit{critical} class and $\epsilon_1^\star = 10^{-3}$ for the \textit{non-critical} class. Furthermore we fix the code rates to be $R_0=1/4$ and $R_1=1/2$.

\begin{figure}[t]
	\centering
	\begin{tikzpicture}[scale=1,every node/.style={scale=0.85}]

\begin{axis}
[
width=\columnwidth,
xmin=50,
xmax=270,
xlabel={Blocklength $n$ [bits]},
ymin=-2.0,
ymax=3.0,
grid=both,
xtick align=outside,
    xtick pos=bottom,
    ytick pos=left,
    xtick align=outside,
    ytick align=outside,
ylabel={$E_s/N_0$ [dB]},
grid style={line width=.1pt, draw=gray!10},
    major grid style={line width=.2pt,draw=gray!50},
    minor tick num=4,
ytick={-4.0,-3.5,-3.0,-2.5,-2.0,-1.5,-1.0,-0.5,0.0,0.5,1.0,1.5,2.0,2.5,3.0},
smooth
]
\footnotesize

\addplot[line width=0.5,color = black, mark=none, mark options={solid}]
table[x=n ,y=EsN0dB, col sep=space,trim cells=true] {figures/NA_Bound_Draper.txt};

\draw[latex-] (axis cs: 160,-0.65) -- (axis cs: 180, 0.1);
\node[align=center,scale = 0.8] at (axis cs: 185, 0.2) {Benchmark \cite{Shkel15}};



\addplot[only marks, mark= +, mark size = 2, color = red] coordinates {(128, 0.0418)};\label{plot:SCL128Random}

\addplot[only marks, mark= star, mark size = 2, color = blue] coordinates {(128, 1.1)};\label{plot:CC_v6_LRT_128}

\addplot[only marks, mark= o, mark size = 2, color = red] coordinates {(128, 1.1)};\label{plot:CC_v6_GLRT_128}

\addplot[only marks, mark= star, mark size = 2, color = blue] coordinates {(128, 0.7)};\label{plot:CC_v8_LRT_128}

\addplot[only marks, mark= o, mark size = 2, color = red] coordinates {(128, 0.7)};\label{plot:CC_v8_GLRT_128}

\addplot[only marks, mark= star, mark size = 2, color = blue] coordinates {(128, 0.4)};\label{plot:CC_v10_LRT_128}

\addplot[only marks, mark= o, mark size = 2, color = red] coordinates {(128, 0.4)};\label{plot:CC_v10_GLRT_128}



\addplot[only marks, mark= +, mark size = 2, color = red] coordinates {(256,-0.5)};\label{plot:SCL256Random} 

\addplot[only marks, mark= star, mark size = 2, color = blue] coordinates {(256, 1.1)};\label{plot:CC_v6_LRT_256}

\addplot[only marks, mark= o, mark size = 2, color = red] coordinates {(256, 1.1)};\label{plot:CC_v6_GLRT_256}

\addplot[only marks, mark= star, mark size = 2, color = blue] coordinates {(256, 0.6)};\label{plot:CC_v8_LRT_256}

\addplot[only marks, mark= o, mark size = 2, color = red] coordinates {(256, 0.6)};\label{plot:CC_v8_GLRT_256}

\addplot[only marks, mark= star, mark size = 2, color = blue] coordinates {(256, 0.2)};\label{plot:CC_v10_LRT_256}

\addplot[only marks, mark= o, mark size = 2, color = red] coordinates {(256, 0.2)};\label{plot:CC_v10_GLRT_256}


\draw[fill = white, opacity = 0.7, draw = none] (axis cs: 170, 2.5) -- (axis cs: 264, 2.5) -- (axis cs: 264, 1.5) -- (axis cs: 170, 1.5); 

\node[] at (axis cs: 175,2.3) {\ref{plot:SCL128Random}};
\node[anchor=west] at (axis cs: 180,2.3) {Polar code, \textcolor{red}{ALRT}};

\node[] at (axis cs: 175,2.0) {\ref{plot:CC_v10_GLRT_128}};
\node[anchor=west] at (axis cs: 180,2.0) {Convolutional code, \textcolor{red}{ALRT} };

\node[] at (axis cs: 175,1.7) {\ref{plot:CC_v10_LRT_128}};
\node[anchor=west] at (axis cs: 180,1.7) {Convolutional code,  \textcolor{blue}{LRT}};


\node[align=center,scale = 0.8] at (axis cs: 114, 1.1) {$\nu = 6$};
\node[align=center,scale = 0.8] at (axis cs: 114, 0.7) {$\nu = 8$};
\node[align=center,scale = 0.8] at (axis cs: 113, 0.4) {$\nu = 10$};

\node[align=center,scale = 0.8] at (axis cs: 242, 1.1) {$\nu = 6$};
\node[align=center,scale = 0.8] at (axis cs: 242, 0.6) {$\nu = 8$};
\node[align=center,scale = 0.8] at (axis cs: 241, 0.2) {$\nu = 10$};

\end{axis}
\end{tikzpicture}%
	\caption{Minimum $E_s/N_0$ as a function of blocklength $n$ to achieve $(\pOne^{\star},\pTwo^{\star})=\left(10^{-3},10^{-5}\right)$ for $2$-UMP codes with rates $R_0 = 1/2$ and $R_1 = 1/4$. \ac{SCL} decoder is used for \ac{CA}-polar codes with list size $L=32$. For \acp{ZTCC} the LRT and the ALRT tests are compared for different encoding memories $\nu$.}  
	\label{fig:MinEsN0dB}
\end{figure}
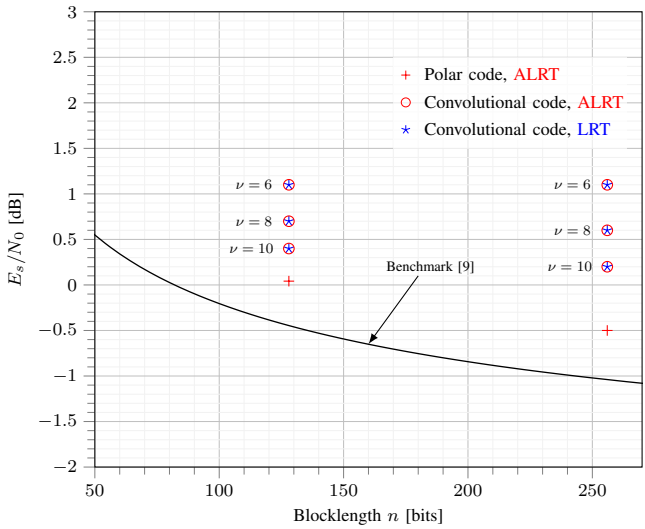

In Fig.~\ref{fig:MinEsN0dB}, we depict the minimum $(E_s / N_0)^\star$ in dB for the \ac{CA} polar and convolutional code construction as a function of the blocklengths $n=128$ and $n=256$. For the \ac{CA} polar codes, we follow the $5$G NR standard~\cite{3GPP21}, using the $11$-bit \ac{CRC} with polynomial $\mathtt{0xE21}$ and the set of information and frozen bits accordingly to the $5$G NR specification. At the decoder we employ \ac{SCL} decoding with list size $L = 32$. Regarding convolutional codes, the generators used in the simulations and expressed in octal notation are $G_0=[117,127,155,171]$ and $G_1=[133,171]$ when $\nu=6$, $G_0=[473,513,671,756]$ and $G_1=[515,677]$ when $\nu=8$ and $G_0=[2565,2747,3311,3273]$ and $G_1=[3645,2671]$ when $\nu=10$. For convolutional codes, interestingly, we observe that the \ac{LRT} and \ac{ALRT} exhibit equivalent performance. Moreover, a negligible performance gain is observed when extending the blocklength from $n=128$ to $n=256$. Polar codes approach the finite-length benchmark of (2). The distance from the benchmark is approximately $0.54~\dB$ for both blocklengths of $n=128$ and $0.62~\dB$ for $n=256$. 

Table~\ref{tab:maxRate} reports the maximum code rate pairs $(R_0^\star, R_1^\star)$ for the target error probabilities $(\epsilon_0^\star, \epsilon_1^\star) = (10^{-4}, 10^{-2})$ at $E_s/N_0 = -3$ dB. For a fixed blocklength $n$, $(R_0^\star, R_1^\star)$ corresponds to the largest rates that satisfy the error constraints $(\pNull^\star,\pOne^\star)$. We compare our $2$-\ac{UMP} coset polar code construction with the polar coding approach using the \ac{BBT} method \cite{Xiao23}. In~\cite{Xiao23}, the proposed construction superimposes a class indicator vector to generate a coset code on a rate-matched \ac{BBT}. For the decoding process, \cite{Xiao23} tests all possible classes to create a list of estimated codewords. Within this list, a codeword is selected that maximizes the likelihood with an offset. Our $2$-\ac{UMP} \ac{CA} polar codes use the $6$-bit \ac{CRC} of the $5$G NR standard~\cite{3GPP21} with polynomial $\mathtt{0x61}$ and \ac{SCL} decoding with list size $L=32$. From Table~\ref{tab:maxRate}, it seems that our two-step decoder achieves similar rates w.r.t. the \ac{BBT} construction for both $n=128$ and $n=256$. This suggests that our two-step decoding strategy, where each step corresponds to an optimization problem, provides simple yet robust rules while delivering also comparable gains in terms of achievable rates, without the need of special code constructions like in \cite{Xiao23}.

\bgroup
\def\arraystretch{1.1}
\begin{table}[h]
\centering
\caption{Maximum achievable rates $(R_0^\star,R_1^\star)$ for our coset \ac{CA} polar construction and the \acs{BBT} approach~\cite{Xiao23} for $E_s / N_0 = -3 \dB$  and $(\pNull^\star,\pOne^\star) = (10^{-2},10^{-4})$.}
\label{tab:maxRate}
\begin{tabular}{l|cc|cc}
 & \multicolumn{2}{c|}{$n=128$ [bits]} & \multicolumn{2}{c}{$n=256$ [bits]} \\
\hline
\ac{CA} polar & $R_0^\star \approx 0.23$ & $R_1^\star \approx 0.30$ & $R_0^\star \approx 0.28 $ & $R_1^\star \approx$ 0.35\\
\hline 
\acs{BBT} \cite{Xiao23} & $R_0^\star \approx 0.25$ & $R_1^\star \approx 0.30$ & $R_0^\star \approx 0.28$ & $R_1^\star \approx 0.35$ \\
\end{tabular}
\vspace{-2mm}
\end{table}
\egroup

\section{Conclusions}\label{sec:conclusion}


In this paper, we investigated the design and decoding of codes in the unequal message protection framework for short-blocklength applications, focusing on coset-based constructions for polar and convolutional codes. By adopting the coset construction, we remove the need for explicit preambles, which typically incur important rate penalties in small-$n$ scenarios. We derived formal criteria to ensure disjointness between message classes and proposed a two-step decoding architecture, where each step follows an optimization problem. 

Numerical results confirm that our construction closely tracks finite-length information theoretic benchmarks, with CA-polar codes exhibiting a near-constant $0.5$ dB gap from the benchmark. Furthermore, our approach demonstrates comparable performance w.r.t. other existing methods, without requiring code design optimizations. These findings highlight the potential of coset-based UMP codes combined with the proposed two-step decoding strategy.

\balance
\bibliography{IEEEabrv,coding}

\end{document}